\documentclass{aa}
\def\lsim{\;\raise0.3ex\hbox{$<$\kern-0.75em\raise-1.1ex\hbox{$\sim$}}\;}
\def\gsim{\;\raise0.3ex\hbox{$>$\kern-0.75em\raise-1.1ex\hbox{$\sim$}}\;}

\begin{document}

\title{Superbubbles and Energetic Particles in the Galaxy}
\subtitle{I: Collective effects of particle acceleration}

\author{E. Parizot\inst{1} \and A. Marcowith\inst{2} \and E. van der Swaluw\inst{3,4} \and A. M. Bykov\inst{5} \and V. Tatischeff\inst{6}}

\institute{Institut de Physique Nucl\'eaire d'Orsay, IN2P3-CNRS/Universit\'e Paris-Sud, 91406 Orsay Cedex, France \and Centre d'\'Etude Spatiale des Rayonnements, 9 av. du Colonel
Roche, 31028 Toulouse Cedex 4, France \and Dublin Institute for Advanced Studies, 5 Merrion Square, Dublin 2, Ireland \and FOM-Institute for Plasma Physics Rijnhuizen, P.O. Box 1207, 3430 BE Nieuwegein, The Netherlands \and A.F.~Ioffe Institute for Physics and Technology, St.~Petersburg, Russia, 194021 \and Centre de Spectrom\'etrie Nucl\'eaire et de Spectrom\'etrie de Masse, IN2P3-CNRS, 91405 Orsay, France}

\date{Received date; accepted date}

\authorrunning{Parizot et al.} \titlerunning{collective effects of CR acceleration in SBs}

\abstract{Observations indicate that most massive stars in the Galaxy appear in groups, called OB associations, where their strong wind activity generates large structures known as superbubbles, inside which the subsequent supernovae (SNe) explode, in tight space and time correlation.  Acknowledging this fact, we investigate four main questions: 1) does the clustering of massive stars and SN explosions influence the particle acceleration process usually associated with SNe, and induce collective effects which would not manifest around isolated supernova remnants?; 2) does it make a difference for the general phenomenology of Galactic Cosmic Rays (GCRs), notably for their energy spectrum and composition?; 3) can this help alleviate some of the problems encountered within the standard GCR source model?; and 4) Is the link between superbubbles and energetic particles supported by observational data, and can it be further tested and constrained?  We argue for a positive answer to all these questions.  Theoretical, phenomenological and observational aspects are treated in separate papers.  Here, we discuss the interaction of massive stellar winds and SN shocks inside superbubbles and indicate how this leads to specific acceleration effects.  We also show that due to the high SN explosion rate and low diffusion coefficient, low-energy particles experience repeated shock acceleration inside superbubbles.
}

\maketitle

\section{Introduction}
\label{sec:intro}

Galactic cosmic-rays (GCRs) are believed to be powered by the mechanical energy of supernova (SN) explosions in the interstellar medium (ISM).  A number of arguments support this hypothesis.  First, global energetics: the total power of Galactic SNe is compatible with the power needed to maintain the observed density of GCRs throughout the Galaxy (and part of the halo).  Second, chemical composition: the overall CR source composition appears to be compatible with particle acceleration out of the general ISM, with some variations probably related to ionization potential or volatility selection effects, as expected from consistent theoretical considerations (Meyer et al.  1997, Ellison et al.  1997).  Third, energy spectrum: the observed, single power-law CR spectrum up to $\sim 3\times 10^{15}$~eV is compatible with a \textit{universal} power-law source spectrum resulting from particle acceleration at supernova remnant (SNR) shocks, corrected for propagation effects (which manifest as a power-law dependence on energy of the CR confinement time in the Galaxy).  This makes it possible for the contributions of each individual source to add up in a simple way, without producing any structure in the spectrum.  The non-universality of the power-law index was a major problem of the initial Fermi model for CR origin.

For the above reasons, and also because we know from the observed synchrotron emission that relativistic electrons \textit{are} present near the shocks of SNRs, the standard scenario of CR origin in the Galaxy involves the acceleration of (part of) the material swept up by the forward shock of SNRs.  Moreover, particle acceleration at collisionless shocks is believed to be reasonably well understood (at least in the test-particle approximation), as a good agreement is reached between theory, numerical simulations and direct observation at interplanetary shocks (Cliver, 2000; Li et al., 2003).

Nevertheless, this standard scenario suffers from a number of persistent problems, and some important questions remain to be answered, as reviewed in the accompanying paper (Paper~II). In particular, the standard model predicts an energy spectrum which seems too hard, a gradient of the CR distribution as a function of Galactocentric radius which seems to steep, and a CR composition which seems to poor in massive stellar ejecta. In addition, the maximum energy of the particles accelerated in SNRs is too low to account for the observed continuous CR spectrum up to the so-called \emph{ankle}, around $3\times 10^{18}$~eV.

One way to cope with these problems is to look for an improved analysis of some aspects of the model, e.g. concerning the propagation of CRs in the ISM (e.g. Ptuskin, 2001), and/or the transport of particles inside the remnant, during acceleration (e.g. Kirk et al. 1996), notably through a more detailed treatment of the magnetic field structure around the shock (e.g. Jokipii, 1987).  Non linear effects have also been taken into account, to improve on the test-particle treatment of diffusive shock acceleration (e.g. Ellison, 2001; Malkov and Drury, 2001).  However, none of the refined or improved models has been successful (up to now) in solving the problems of the standard GCR source scenario.  In particular, the maximum energy problem remains critical and it seems unlikely that one could solve it without bringing in new ideas.  It should be noted, in particular, that allowing for large fluctuations of the magnetic field around the SN shock (up to 100 times the ambient field or more; Lucek and Bell, 2000; Bell and Luceck, 2001; Berezhko et al., 2003; Ptuskin and Zirakashvili, 2003) does help to reach energies around or even above the knee ($3\times 10^{15}$~eV), but the ankle energy seems to remain inaccessible, even for Iron nuclei, because of the intrinsic, non-linear damping of the required MHD waves and the small amount of time available for highly efficient acceleration (Ptuskin and Zirakachvili, 2003).

An other possibility is to look for alternative scenarios, with radically different models, in the line of what has been proposed notably for ultra-high-energy CRs, e.g. involving neutron stars (de Gouveia dal Pino and Lazarian, 2000), gamma-ray bursts (Waxman, 1995; Vietri, 1995; Pelletier, 1999), active galactic nuclei (Rachen and Biermann 1993; Biermann 1997 and ref.  therein; Henri et al., 1999).
    
In the current study, we shall follow a more conservative approach and keep the main assumption that GCRs are related to SN explosions in the Galaxy.  In the standard scenario for CR origin, the acceleration of particles implicitly occurs at the shocks of \textit{isolated} SNRs.  These SNRs are familiar and their dynamical evolution in a roughly homogeneous ISM is well understood, as a succession of free expansion, adiabatic Sedov-like and radiative snow-plow phases (e.g. Woltjer, 1972).  They have been extensively studied through multi-wavelength analysis, enabling thorough and instructive comparison of the models (dynamics, particle acceleration, radiative transfer, etc.)  with the observational data.  However, isolated SNe represent only a fraction of all stellar explosions in a galaxy, since most SN progenitors are observed in OB associations and thus SN explosions are strongly correlated in space and time.  Therefore, besides the problems of the standard scenario for CR origin, it is natural to investigate the influence of superbubbles (produced by the joint stellar activity of an OB association) on the acceleration processes (Bykov and Toptygin, 1982, 1988; Bykov and Fleishman, 1992; Bykov, 2001) and their role in the production of GCRs (Bykov and Toptygin 1990, 2001; Higdon et al., 1998; Parizot, 2001).

In this series of papers, we shall address the following questions: 1) Does the clustering of massive stars and SN explosions influence the particle acceleration process and induce collective effects which would not manifest in isolated SNRs?; 2) What difference does it make for the GCRs, notably for their energy spectrum and composition?; 3) Can this help alleviate some of the problems encountered within the standard GCR source model?; 4) Is the link between superbubbles and energetic particles supported by observational data, and how can it be further tested and constrained?

The first paper will concentrate on theoretical issues related to collective acceleration effects. The second one will address in greater detail the question of cosmic-ray origin, and investigate the phenomenological aspects of the proposed \textit{superbubble model}. And the third paper will be devoted to the direct and indirect observational counterparts of superbubbles.

\section{OB associations and superbubbles}

\subsection{Distribution of massive stars and SNe in the Galaxy}
\label{sec:DistribMassiveStars}

Most massive stars are formed in groups by the collapse of giant molecular clouds (GMC), with typical sizes of 10 to 30 pc (de Geus, 1991).  Because of their short lifetimes (3 to 20 Myr; e.g. Schaller et al., 1992), these SN progenitors do not have time to acquire large dispersion velocities, and observations confirm typical values of 4--6~km/s (Blaauw, 1991; Mel'nik and Efremov, 1995).  As a consequence, despite the fact that they do not form gravitationally-bound groups, they remain concentrated during their whole life and explode close to their birth place, in relatively compact regions.  This is the reason why massive stars (i.e. O and B stars) are found in associations in the Galaxy.
    
 Although identifying the membership of a given OB association is not an easy task from the observational point of view, it is reliably estimated that between 60\% (Garmany, 1994) and 95\% (Higdon et al., 1998) of all OB stars belong to such associations, which contain up to several tens of OB stars (say between 10 and 100), within regions of radius $R_{\mathrm{OB}}\sim 35\,\mathrm{pc}$ (Garmany, 1994; Bresolin et al., 1999; Pietrzy\`nski, et al., 2001 and references therein).
    
 For evenly distributed stars, the mean distance between two closest neighbours can be evaluated as $D_{\star} \simeq (4\pi R_{\mathrm{OB}}^{3}/3 N_{\mathrm{OB}})^{1/3}$, so that each star can be considered as occupying an individual spherical volume of radius $R_{\star} \simeq D_{\star}/2$, with typical value: \begin{equation} R_{\star} \simeq (6\,\mathrm{pc}) \left(\frac{R_{\mathrm{OB}}}{35\,\mathrm{pc}}\right) \left(\frac{N_{\mathrm{OB}}}{100}\right)^{-1/3}.     \label{eq:Rstar} \end{equation} It should be noted, however, that most OB associations show substructures, referred to as OB subgroups, as a consequence of a complex, hierarchical process of star formation inside GMCs (e.g. de Geus et al., 1989, for Sco-Cen OB2; Brown et al., 1994, for Orion).  These subgroups have smaller numbers, but higher densities of massive stars, with mean distances between closest neighbours sometimes much smaller than the above average value. In the 30 Doradus complex in the LMC, Walborn et al.  (1999) find compact subgroups of massive stars containing typically ten OB stars within few pc, a trend which is confirmed by recent Chandra arcsecond observations (Townsley et al., 2003).  In particular, the star cluster R136 in 30 Doradus contains 9 O stars within 3.4 pc, corresponding to $R_{\star}\sim 1.3$~pc!
    
 Finally, let us recall that roughly 90\% of the SNe exploding in our Galaxy are so-called \textit{core-collapse SNe}, i.e. originating from massive stars (van den Bergh and McClure, 1994; Ferri\`ere, 2001).  Combining that number with the fraction of OB stars in associations, one should expect that the majority, and possibly up to 85\% of the Galactic SNe explode in compact regions around OB associations.  This implies that the energy which is thought to power CR acceleration is not released randomly in the ISM, to form the well-known, independent and isolated SNRs, but mostly on relatively short timescales ($\sim 20$~Myr) in concentrated regions of no more than a few tens of pc.  This energy is released in the form of stellar winds and SN explosions which interact with each other to produce the large Galactic structures known as superbubbles (SBs), as discussed below.

\subsection{The formation of a `super wind bubble'}
\label{sec:SuperWindBubble}

Let us now indicate how the SBs are produced from the collective activity of massive stars in OB associations.  An important characteristic of such stars is that they experience strong winds during most of their lifetime.  The mass-loss rate and the wind velocity -- and thus the wind power -- are not constant during stellar evolution (e.g. Schaller et al., 1992; Meynet et al., 1994), but the total wind energy, integrated over a massive star's lifetime, amounts typically to $10^{51}$~erg and is therefore comparable to the final SN explosion energy itself.  When considering the energy output of OB stars in the Galaxy, one thus has to include the contribution of the winds, which can roughly double the energy imparted to cosmic rays if the wind energy can somehow be used to accelerate particles.  As we discuss below, superbubbles may be an environment where the SN energy \textit{and} the stellar wind energy can be efficiently converted into cosmic rays.
    
 Massive stellar winds also have a strong influence on the dynamics of the ISM around OB associations.  Let us first assume that the individual stellar wind bubbles do not interact with each other, and that each star blows a steady wind with a typical average power of $L_{\mathrm{wind}} \simeq 3\times 10^{36}$~erg/s, in a homogeneous medium of density $n \simeq 10^{2}\,\mathrm{part/cm}^{3}$ (a typical average density for the parent GMC).  According to standard wind bubble theory (Weaver et al., 1977), the radius of the external shock in the semi-adiabatic phase is given as a function of $t_{\mathrm{Myr}}$, the time in Myr, by: \begin{equation} R_{\mathrm{ext}} \simeq (13\,\,\mathrm{pc}) \,t_{\mathrm{Myr}}^{3/5} \left(\frac{L_{\mathrm{wind}}}{3\,10^{36}\, \mathrm{erg/s}}\right)^{1/5} \left(\frac{n}{10^{2}\,\mathrm{cm}^{-3}}\right)^{-1/5}.     \label{eq:Rext} \end{equation}
    
 This is significantly larger than the mean half-distance between massive stars in the association, $R_{\star}$ (Eq.~\ref{eq:Rstar}), so that the individual wind bubbles actually collide and merge during the first million year of stellar activity.  The result is a large, collective bubble expanding almost spherically (in a homogeneous medium) around the whole OB association, similarly to a standard wind bubble that would simply be powered by the sum of the mechanical luminosity of each individual wind.

It is thus found from this simple picture that the SBs around OB associations should actually form before the first SN explosion, from the combined activity of stellar winds.

\subsection{Inhomogeneities and clumps}
\label{sec:clumps}

In practice, molecular clouds cannot be considered as homogeneous: they contain many \textit{clumps} with a variety of densities, typically ranging from $10^{3}$ to $10^{6}\,\mathrm{cm}^{-3}$, or even much more in the localized regions where stars will eventually form.  The effect of such clumps on the evolution of the wind bubbles and the collective superbubble will be analyzed in more detail in a forthcoming paper.  Here, we simply note that high-density clumps around massive stars cannot be swept-up by the winds and integrated into the expanding shells.

A rough estimate can also be obtained as follows.  In order for the clump not to be carried away by the wind, an approximate condition is that its column density be larger than that of the wind shell when the winds collide and the superbubble forms, i.e. roughly when the wind shell radius is larger than $R_{\star}$. Comparing the clump column density, $\sim \frac{4}{3}n_{\mathrm{cl}}R_{\mathrm{cl}}$, with that of the largest individual wind shell, $\sim \frac{1}{3}n_{\mathrm{GMC}}R_{\star}$, one obtains a condition for GMC clumps to remain inside the growing SB around an OB association:
\begin{equation}
n_{\mathrm{cl}} \ga (1.5\,10^{3}\,\mathrm{cm}^{-3}) \left(\frac{R_{\mathrm{cl}}}{10^{-1}\,\mathrm{pc}}\right)^{-1} \left(\frac{n_{\mathrm{GMC}}}{10^{2}\,\mathrm{cm}^{-3}}\right) \left(\frac{R_{\star}}{6\,\mathrm{pc}}\right)
\label{eq:nCl}
\end{equation}
In other words, reasonably dense clumps, unless they are insignificantly small, will remain trapped inside the SB. Note that condition (\ref{eq:nCl}) could actually be made less severe by taking into account the inertia of the clump, or if one prefers, the fact that the wind shell is actually less massive when it encounters the clumps at stellocentric distances smaller than $R_{\star}$.

Apart from wind sweeping, the intense ultra-violet radiation accompanying the OB association stellar activity could also destroy clumps.  The extreme-UV photons ionize the gas surrounding massive stars, forming HII regions, and the far-UV photons also dissociate molecular gas beyond the HII region (Hollenbach and Tielens, 1999).  This results in a rapid homogenization of the less dense regions, with cloud densities around 1-10 $\mathrm{cm^{-3}}$.  However, despite the continuous erosion of the molecular gas, denser molecular globules survive in the HII region and are slowly advected with the ionized gas (Bertoldi and McKee, 1990).  The typical size of such globules are $\sim 0.01-0.1 \mathrm{pc}$, and up to 1~pc.

From the hydrodynamical point of view, when a shock front hits an overdense clump, it generates a reflected shock, in addition to the transmitted shock propagating more slowly inside the clump. As the shock fronts progress around them, the clumps of highest density find themselves engulfed in the bubbles, and the SB forms around them through the successive merging of individual bubbles. Some evaporation of the shocked material will occur, but the interior of the early superbubble should remain very clumpy, with localized high density contrasts.  Parts of the individual bubble shells can also be trapped inside the SB, with typical sizes on the parsec scale.  Later on, when shock fronts from a new wind phase (e.g. Wolf-Rayet) or a new SN propagate inside the SB, the same mechanism recurs, with the denser globules surviving and producing reflected shocks (see e.g. the numerical work of Poludnenko et al., 2002).

During the process of SB formation and growth, a substantial fraction of the energy contained in the parts of the shells which encounter the clumps (or which encounter each other) will be transferred to secondary shocks and turbulence.  Shock-clump interactions should also lead to efficient MHD waves generation, especially since the dense clumps in massive star formation sites are known to be highly magnetized.  Zeeman effect measurements, which probe the line-of-sight field, show magnetic field values as high as a few to a few tens of milligauss (Sarma et al., 2002). Less dense clumps, with densities $10^{3}$ -- $10^{6}\,\mathrm{cm}^{-3}$, also have large magnetic fields of a few tens of $\mu$G (Crutcher, 1999).

It should also be noted that, in addition to the generation of MHD turbulence through the coupling with large scale hydrodynamic motions with velocities close to the Alfv\'en speed (cascading to smaller scales down to the gyroradius of thermal protons; e.g. Goldreich and Sridhar, 1997), relativistic particles can generate or amplify MHD waves notably through streaming instability (see e.g. Bykov et al., 2000).

\section{Stellar winds inside superbubbles}

Once the superbubble is formed by the merging of the wind bubbles, its interior consists of a hot, low-density medium (apart from the above-mentioned clumps), where the shocked wind material of all the stars match together subsonically.  Closer to the stars, however, is a region of unshocked wind material blowing roughly spherically at supersonic velocities, producing a strong wind termination shock.  It is instructing to calculate the typical radius, $R_{\mathrm{term}}$, of such shocks.  It is obtained by equating the ram pressure of the wind, $P_{\mathrm{ram}} = \rho_{\mathrm{w}} V_{\mathrm{w}}^{2} = \dot{M}_{\mathrm{w}}V_{\mathrm{w}}/4\pi r^{2} = L_{\mathrm{w}}/2\pi V_{\mathrm{w}}r^{2}$, and the thermal pressure in the SB interior, $P_{\mathrm{SB}}$, which depends on the dynamical evolution of the superbubble, and is a decreasing function of time.

\subsection{Typical physical conditions in SB interiors}

The theory of SB evolution has given rise to a lot of work (Mac Low and McCray, 1988; Tomisaka, 1990; Koo and McKee, 1992; Shull and Saken, 1995; Korpi et al., 1999; Silich and Franco, 1999).  They all are based on the standard wind bubble theory (Weaver et al., 1977), but differ in the treatment of some aspects of the SB dynamics.  Both analytical and numerical studies have been provided, taking into account external magnetic fields, density gradients, inhomogeneous environments, thermal conduction, shell evaporation, mass loading from internal clouds, etc.  Here, we shall not go into any particular detail, as we are only interested in the typical values of basic physical quantities inside the SB. To this aim, we follow Mac Low and McCray (1988) and assume that the SB is expanding in a homogeneous medium of density $n_{0}$, powered by the activity of an OB association providing a constant mechanical luminosity, $L_{\mathrm{OB}}$.

The energy release inside the SB is not continuous and experiences strong peaks when an OB star enters the Wolf-Rayet stage or when a new SN explodes.  However, it can be shown that for sufficiently evolved SBs (after a few Myr, say) the variations of the driving power are smoothed out, as the shells of individual SN shocks become subsonic before they reach the supershell (except of course for SNe exploding particularly close to it), and their energy is turned into internal energy before it can have direct influence on the supershell dynamics (see below, and Mac Low and McCray, 1988).  The SB interior thus acts as a buffer which absorbs the rapid variations of the input power.

In addition, star formation is a sequential process in GMCs, and massive stars have a whole range of lifetimes (see above).  It is therefore legitimate, as a first approximation, to assume that the energy release is indeed roughly constant inside the SB, which allows us to treat the whole SB as a very large wind bubble, with ``superwind'' power $L_{\mathrm{OB}} = L_{\mathrm{OB},38}\times 10^{38}\,$~erg/s.  The supershell then refers to the large shell of cool ($T \sim 10^2$~K) and dense ($n_{sh} \sim 100 \,\mathrm{cm}^{-3}$) gas surrounding the whole SB, powered by both winds and SNe (individual, expanding shells can be found inside the SB).

With the above assumptions, Mac Low and McCray (1988) follow Weaver et al. (1977) to find the temperature and density inside the SB:
\begin{equation}
T_{\mathrm{SB}} \simeq (3.5\,10^{6}\,\mathrm{K})\, L_{\mathrm{OB},38}^{8/35}\,n_{0}^{2/35}\,t_{7}^{-6/35}f(x)\,,
\label{eq:TSB}
\end{equation}
and
\begin{equation}
n_{\mathrm{SB}} \simeq (4.0\,10^{-3}\,\mathrm{cm}^{-3})\, L_{\mathrm{OB},38}^{6/35}\,n_{0}^{19/35}\,t_{7}^{-22/35}f(x)^{-1}\,,
\label{eq:nSB}
\end{equation}
where $n_{0}$ is the external ISM density in $\mathrm{cm}^{-3}$, $t_{7}$ is the age of the SB in units of $10^{7}$~yr, $x = r/R_{\mathrm{SB}}$ is the relative distance from the SB center, and $f(x) = (1-x)^{2/5}$ is a function giving the temperature and density gradient inside the SB, assuming that the energy is injected at $x = 0$.  Even though such a simplification is not realistic, we are only interested in the resulting estimate of the internal pressure, i.e. the product $P_{\mathrm{SB}} = \mu n_{\mathrm{SB}}k_{\mathrm{B}}T_{\mathrm{SB}}$, which is independent of $f(x)$.  The average pressure is indeed constant inside the SB, as the sound crossing time is lower than the SB dynamical time.  Here we assume a particle multiplicity $\mu \simeq 2.3$, taking into account the contribution of the electrons to the pressure (the SB interior is here assumed to be fully ionized, with solar abundances).

The SB internal pressure thus reads:
\begin{equation}
P_{\mathrm{SB}} \simeq (4.3\,10^{-12}\,\,\mathrm{dyne\,\,cm}^{-2})\, L_{\mathrm{OB},38}^{2/5}\,n_{0}^{3/5}\,t_{7}^{-4/5}.
\label{eq:PSB}
\end{equation}

\subsection{Wind-wind interaction}

\begin{table*}
\caption[]{Typical stellar wind parameters in the three main phases of the evolution of a $35\,M_{\odot}$ star (main sequence, red supergiant, Wolf-Rayet) and a $60\,M_{\odot}$ star (main sequence, luminous blue variable, Wolf-Rayet): duration of the phase, stellar mass loss rate, final wind velocity, wind mechanical luminosity, total energy of the wind in the corresponding phase, wind overlap ratio (see text). The latter scales as $t_{7}^{2/5}L_{\mathrm{OB},38}^{-1/5}\,n_{0}^{-3/10} (R_{\mathrm{OB}}/35\,\mathrm{pc})(N_{\mathrm{OB}}/100)^{-1/3}$.}
\label{WindParameters}
$$
\begin{array}{lcccccc}
\hline
\noalign{\smallskip}

\mathrm{stellar\,model} & \mathrm{duration} & \dot{M}_{\mathrm{w}} & V_{\mathrm{w}} & L_{\mathrm{w}} & E_{\mathrm{w}} & R_{\mathrm{term}}/R_{\star} \\
\mathrm{mass/phase} & \mathrm{Myr} & 10^{-5}\,M_{\odot}/\mathrm{yr} & 10^{3}\,\mathrm{km/s} & 10^{37}\,\mathrm{erg/s} & 10^{50}\,\mathrm{erg} & (t_{\mathrm{OB}} = 10\,\mathrm{Myr}) \\
\noalign{\smallskip}
\hline
\noalign{\smallskip}
35\,\mathrm{M}_{\odot}/\mathrm{MS}  & 4.2 & 0.06 & 3.1   & 0.2   & 2.6   & 0.85 \\
35\,\mathrm{M}_{\odot}/\mathrm{RSG} & 0.2 & 9.0  & 0.075 & 0.017 & 0.011 & 1.6  \\
35\,\mathrm{M}_{\odot}/\mathrm{WR}  & 0.2 & 2.2  & 2.0   & 2.9   & 1.8   & 4.0  \\

60\,\mathrm{M}_{\odot}/\mathrm{MS}  & 3.4   & 0.94 & 3.1  & 3.1  & 33.  & 3.3 \\
60\,\mathrm{M}_{\odot}/\mathrm{LBV} & 0.012 & 65.  & 0.4  & 3.4  & 0.13 & 9.7 \\
60\,\mathrm{M}_{\odot}/\mathrm{WR}  & 0.6   & 2.7  & 2.5  & 5.6  & 11.  & 5.0 \\
\noalign{\smallskip}
\hline
\end{array}
$$
\end{table*}

We can now find the radius of a stellar wind termination shock, around a given O or B star with a wind power $L_{\mathrm{w}} = L_{\mathrm{w},37} \times 10^{37}$~erg/s, and a wind velocity $V_{\mathrm{w}} = V_{\mathrm{w},3}\times 10^{3}$~km/s, by equating $P_{\mathrm{SB}}$ from Eq.~(\ref{eq:PSB}) with the wind ram pressure, $P_{\mathrm{ram}}$, which depends on the stellocentric distance $r_{\mathrm{pc}}$ (in parsecs) as:
\begin{equation}
P_{\mathrm{ram}} \simeq (1.7\,10^{-9}\,\,\mathrm{dyne\,\,cm}^{-2})\, L_{\mathrm{w},37}\,V_{\mathrm{w},3}^{-1}\,r_{\mathrm{pc}}^{-2}.
\label{eq:PRam}
\end{equation}

One obtains:
\begin{equation}
R_{\mathrm{term}} \simeq (20\,\mathrm{pc})\, L_{\mathrm{w},37}^{1/2}\,V_{\mathrm{w},3}^{-1/2} L_{\mathrm{OB},38}^{-1/5}\,n_{0}^{-3/10}\,t_{7}^{2/5}.
\label{eq:Rterm}
\end{equation}

Interestingly enough, this radius can be larger than $R_{\star}$, given by Eq.~(\ref{eq:Rstar}).  Even in the early stages of SB evolution, say a few million years after the onset of the wind activity, the free wind termination radius, $R_{\mathrm{term}}(t_{7} = 0.3) \ga 12$~pc, is larger than the distance between two stars in a typical OB association, for typical parameters of strong stellar winds.  Table~1 gives typical wind parameters for massive stars of $35\,M_{\odot}$ and $60\,M_{\odot}$, in three different phases of their evolution, as gathered from Garcia-Segura et al.  (1996a,b).  In the last column, we have indicated the value of the `overlap ratio', which we define as $R_{\mathrm{term}}/R_{\star}$, $10^{7}$~years after the beginning of the SB growth.  As can be seen, this ratio is as high as 4 or 5 in the Wolf-Rayet stellar evolutionary stage, and larger than 1 even in the less powerful main-sequence phase, for massive enough stars.  We therefore conclude that direct wind-wind interaction should occur inside superbubbles.

It must be stressed that this interaction is very different from the merging of wind bubbles discussed in Sect.~\ref{sec:SuperWindBubble}.  There, it was the shocked subsonic wind material of two different stars which was put into contact as the wind bubbles expanded.  The shells of swept-up circumstellar material then merged into a larger shell, pushed further ahead in the ISM by the high pressure in the shocked material inside the collective bubble.  Here, we find that due to the low pressure inside the SB (even just a few Myr after its formation), the region containing \textit{unshocked} wind material extends far enough around the star so that it may enter directly into contact with the unshocked wind material of another star.  Of course, since both winds are highly supersonic, the direct contact cannot actually occur, and a termination shock forms ahead of each wind, where the material blown from each star gets shocked and becomes subsonic.  This situation can be described by saying that the winds actually \textit{terminate each other}, instead of being terminated by the surrounding medium.

As a consequence the interaction region has a much higher pressure than the typical pressure of the SB interior, and it is expected (and indeed confirmed by numerical simulations of single colliding winds; e.g. Pittard, 1998;  Walder and Folini, 2000) that the termination shocks are then very instable.  The wind energy contained in the solid angle where the winds terminate each other (which can be close to one for sufficiently evolved SBs, with low internal pressure, or for dense OB subgroups) is then efficiently converted into strong turbulence, and since the material is fully ionized, plasma waves should also rapidly develop and produce a magnetic turbulence with values of the magnetic field close to the equipartition value.  As can be checked from the corresponding wind parameters, this mechanism will be particularly efficient during the post main-sequence evolution of massive stars.  While these stage are relatively short (a fraction of Myr), the integrated power of the wind can be quite high, and comparable to the SN explosion energy, as shown in the last but one column of Table~\ref{WindParameters}.  Wind-wind interaction in SB cores can thus be an important mechanism to feed strong magnetic turbulence with energy.

In this respect, it may be interesting to note that recent Chandra observations (Townsley et al., 2003) have provided X-ray images with high spatial resolution of two HII regions, known to be compact high-mass star forming regions: the Omega Nebula, M 17, and the Rosette Nebula, NGC 2237-2246.  In both cases, Chandra detected a diffuse soft X-ray emission on parsec scales, which is spatially and spectrally distinct from the point source population.  The luminosity of these diffuse emissions are $L_{\mathrm{X}}\simeq 3.4\,10^{33}$ ergs/s and $L_X \simeq 6.2\,10^{32}$ ergs/s respectively, and can be understood if the $\sim 10 \%$ of the OB stars wind energy is converted into shocks.  As there is no strong evidence of any SN explosion that could contribute to this emission, it is believed that O star wind-wind interactions or the interaction of wind termination shocks with dense molecular clumps are probably responsible for the dissipation of the wind energy into X-rays.

\subsection{Strong turbulence generation in SB core}

The fraction of the wind energy which should participate to the generation of turbulence through the above-mentioned mechanism depends on the wind overlap ratios of the various stars in their different phases.  Since the wind parameters depend on the initial stellar mass, one may expect differences between clusters, the ones containing more massive stars being more active, in the sense of having stronger wind-wind interactions.  For example, the presence of a $60\,\mathrm{M}_{\odot}$ in an OB association can by itself produce an active SB core, since the corresponding wind termination shock radius, $R_{\mathrm{term}}$, in a typical 10~Myr old SB, will be of the order of 20~pc during the MS phase, and even more afterwards, and therefore encompass several neighbouring OB stars (or even a whole subcluster!).  The wind of the other stars, even if they are weaker, will thus be terminated by the most powerful one, leading to strongly fluctuating contact discontinuities and chaotic turbulence generation.  Since the wind velocities are supersonic, a strong turbulence will actually develop, with important intermittency modulated by the changes in the stellar wind phases (Walder and Folini, 2000).

On larger timescales, the generation of turbulence by the OB winds inside the SB core will also be modulated according to the sequence of star formation and the retarded onset of the Wolf-Rayet phase.  The initial mass function will also play an important role in determining the energy conversion efficiency.

While the above estimates have been obtained under the assumption of regularly spaced massive stars (with a spacing of $2R_{\star}$), the effective overlap ratios in the SB core depend on the actual distribution of stars in the OB association.  Any sub-clustering in the association will result in smaller distances between closest neighbours, and OB subgroups with a larger density of stars will have a lower $R_{\star}$ and thus larger overlap ratios.  In the case of R136 in 30 Dor, mentioned in Sect.~\ref{sec:DistribMassiveStars}, the value of $R_{\star}$ is 4 times smaller than our fiducial value, so that massive stars in the WR stage will have termination shocks encompassing the whole subgroup (leading to strong wind-wind interactions), and less massive OB stars will also have overlap ratios larger than 1, possibly even during the main-sequence evolutionary stage.

Therefore, we can expect that, at least in some OB associations (and probably in most of them), a significant fraction of the total wind energy (which can be larger than the SN explosion energy for the most massive stars) should be processed through direct wind-wind interaction in a typical OB association. Efficient particle acceleration should then take place in the resulting strong turbulence and MHD waves, as further discussed below.

In addition, we should keep in mind that, as argued above (see Sect.~\ref{sec:clumps}), the medium around OB stars in a superbubble should be inhomogeneous and contain numerous high density clumps and filaments, inherited from the SB formation process as well as due to previous wind-wind and shock-clumps interactions.  Since the strong wind termination shocks are found to occupy a significant fraction of the SB cores, it is also expected that most of these clumps will be encountered by the supersonic winds, leading to numerous secondary shocks, as well as MHD waves.  The latter will be generated all the more efficiently that the high density clumps inside GMCs are strongly magnetized (cf.  Sect.~\ref{sec:clumps}).

In conclusion, the proximity of the massive stars in the OB association and the low value of the SB internal pressure make it possible for winds to `collide' and terminate each other, imparting a significant fraction of the OB association's wind energy into turbulence and MHD waves, which is further reinforced by the interaction of the supersonic winds with high density clumps and filaments in the SB core.  This is an important feature of efficient acceleration models inside superbubbles, where advantage can be taken of the concomitance of strong stellar activity in a restricted volume: \emph{the collective effect of all OB stars in the association does not come down to the sum of the individual effects of isolated massive stars}.

\section{Supernov\ae~inside superbubbles}

As shown above, the environment in which most SNe explode in the Galaxy is very different from the average ISM which is found around the most studied, isolated SNe.  We now discuss in what respect this can influence the evolution of SNRs.

\subsection{Distortion of the shock front}
\label{sec:distortion}

Bykov (1982) has shown that the propagation of a shock front in a turbulent flow leads to some distortion which can be represented by random relative displacements of individual sections of the front.  This does not destroy the front, however, as a saturated regime is reached where the growth of the distortions is dumped by the propagation of surface waves (and MHD waves in a plasma with $\beta\sim 1$, as in SBs) along the shock front.  In this way, some energy is pulled out of the shock into the (magnetic) turbulence, and strong fluctuations of the electron density are produced, on scales $\ga 10^{13}$~cm (Bykov, 1982).  This can account for the scintillation measurement of background pulsar light (e.g. Rickett 1990).  These distortions and the accompanying electron and magnetic field fluctuations will also influence the diffusive shock acceleration process for particles with gyroradii comparable to the typical amplitude of the shock front perturbations.

\subsection{SNR evolution}

As far as the global behaviour of the SNR is concerned, one should not expect significant deviations from the standard laws describing the evolution of the shock radius and velocity in a homogeneous medium (except if shock quenching occurs, as mentioned below).  It is interesting, however, to scale these laws according to the physical conditions encountered in SB interiors: a lower density will result in a quicker growth of the SNR shell, and a higher temperature in a higher sound speed.

The first stage of SNR evolution corresponds to a free expansion of the ejecta, at a roughly constant velocity scaling like $v_{\mathrm{SN}} = (2E_{\mathrm{SN}}/M_{\mathrm{ej}})^{1/2} \simeq 3.2\,10^{3}\,\mathrm{km/s}\times (E_{51}/M_{10})^{1/2}$, where $E_{51}$ is the explosion energy in units of $10^{51}$~erg and $M_{10}$ is the mass of the ejecta in units of $10\,M_{\odot}$. A transition to a Sedov-like expansion occurs when a mass of roughly $1.6\,M_{\mathrm{ej}}$ is swept-up by the ejecta (McKee and Truelove, 1995). In a medium of density $4\,10^{-3}\,\mathrm{cm}^{-3}$, which corresponds to the case of our typical SB after $10^{7}$~yr of evolution (see Eq.~(\ref{eq:nSB})), this occurs when the SNR reaches a radius $R_{0} \simeq 30\,\mathrm{pc}\times M_{10}^{1/3}\,t_{7}^{0.21}$, i.e. $\sim 1.3\,10^{4}\,\mathrm{years}\times (M_{10}^{5/6}\,t_{7}^{0.21}E_{51}^{-1/2})$ after the explosion (Truelove and McKee, 1999, with parameters $n = 7$ and $s = 0$).

In the Sedov-like phase, the SNR expands almost self-similarly (if we except a small time offset), as from a point explosion, according to:
\begin{equation}
R_{\mathrm{SNR}} \simeq (38\,\mathrm{pc})\, t_{\mathrm{SNR},4}^{2/5}\,t_{\mathrm{SB},7}^{0.126}
\label{eq:RSNR}
\end{equation}
and
\begin{equation}
V_{\mathrm{SNR}} \simeq (1470\,\mathrm{km/s})\, t_{\mathrm{SNR},4}^{-3/5}\,t_{\mathrm{SB},7}^{0.126}
\label{eq:VSNR}
\end{equation}
where we have replaced the ambient gas density by that of the SB interior given by Eq.~(\ref{eq:nSB}) and used the values corresponding to our fiducial SB, and where $t_{\mathrm{SB},7}$ is the age of the SB in units of $10^{7}$~yr and $t_{\mathrm{SNR},4}$ the age of the SNR in units of $10^{4}$~yr (0.126 is an approximate value of the ratio 22/175).  Note also that the dependence on the OB association luminosity and ambient (ISM) density is very low (power indices of 0.034 and 0.11 respectively).

The above equations allow one to calculate the time when the shock becomes subsonic.  Replacing the temperature, $T_{\mathrm{SB}}$, from Eq.~(\ref{eq:TSB}), in the expression of the sound velocity, $c_{\mathrm{s}} = \sqrt{\gamma p/\rho} \simeq \sqrt{\gamma kT/(1.4\,m_{\mathrm{p}})} \simeq 99\,\mathrm{km/s}\times(T/10^{6}\,\mathrm{K})^{1/2}$ for $\gamma = 5/3$, one finds that $V_{\mathrm{SNR}} > c_{\mathrm{s}}$ until
\begin{equation}
t_{\mathrm{sub}} \simeq (3.1\,10^{5}\,\mathrm{yr})\, t_{\mathrm{SB},7}^{37/105}.
\label{eq:tSub}
\end{equation}
By that time, the SNR has reached a radius
\begin{equation}
R_{\mathrm{sub}} \simeq (150\,\mathrm{pc})\, t_{\mathrm{SB},7}^{4/15},
\label{eq:RSub}
\end{equation}
which can be compared to the radius of the superbubble itself (e.g. Mac Low and McCray, 1988):
\begin{equation}
R_{\mathrm{SB}} \simeq (267\,\mathrm{pc})\, L_{\mathrm{OB},38}^{1/5}\,n_{0}^{-1/5}\,t_{7}^{3/5}.
\label{eq:RSB}
\end{equation}

The important point to note is that $R_{\mathrm{sub}} < R_{\mathrm{SB}}$ very early in the evolution of a typical SB (i.e. for the very first SN), and that $R_{\mathrm{sub}}/R_{\mathrm{SB}}$ still decreases as $0.56\,t_{7}^{-1/3}$ as the SB evolves.  In other words, when a SN explodes inside a superbubble, its forward shock will never reach the supershell, unless the explosion site is exceptionally close to it.  The energy of the SNR shell will eventually aliment the SB internal energy, as the expanding shock becomes subsonic and the shock fades into heat and sonic waves.  This justifies the statement made above that the discrete energy releases inside the SB are actually smoothed out and the growth of the SB can be worked out by assuming a continuous driving power.

Another very important information can be derived from the above scaling of the evolution of a SNR in a hot, rarefied medium. Contrary to what occurs for isolated SNe in the ISM, one can show that the shell of a SN exploding inside a SB becomes subsonic \textit{before} becoming radiative.  Indeed, evaluating the cooling time of the shocked gas compared to the age of the SNR, Blondin et al.  (1998) obtained the timescale for the end of the Sedov-like phase and the formation of a radiative shell as $t_{\mathrm{rad}}= 2.9\,10^4\,\mathrm{yr} E_{51}^{4/17} n_{0}^{-9/17}$.  Replacing $n_{0}$ by the SB density and dividing by the time corresponding to the sonic transition, one finds that
\begin{equation}
t_{\mathrm{rad}}/t_{\mathrm{sub}} \simeq 1.7\times t_{7}^{-1/51}
\label{eq:tRadOverTSub}
\end{equation}
is always larger than 1, indicating that the SNR will never become radiative inside the SB. Although this conclusion depends in principle on the parameters of our typical SB, the dependence appears to be very weak, in $L_{\mathrm{OB}}^{0.091}$ (and $n_{0}^{0.29}$).  One may therefore be confident that SN shock waves remain in the Sedov-like phase (and thus keep their initial energy) until they die well inside the SB.

\subsection{Energy balance}

This makes a significant difference when considering particle acceleration efficiency: while a substantial fraction of the SNR energy is radiated away in isolated SNRs, and thus not available for particle acceleration, all the SNR energy is eventually turned into internal energy inside a SB. Now, not only heat is produced in this way, but given the characteristics of the SB interior, with pre-existing turbulence and inhomogeneities (see above), one can expect that the dying SNR shocks will turn a substantial fraction of their energy into additional turbulence, which is an important ingredient of the acceleration mechanism discussed in Sect.~\ref{sec:turbulentAcc}.  Magnetic turbulence will also be produced (or amplified) during the alfv\'enic transition, which should occur before the sonic transition if the Alfv\'en velocity is in fact larger than the sound speed in SBs, as we expect if $B_{\mathrm{SB}}\ga 10\,\mu\mathrm{G}$ (see Sect.~\ref{sec:DSAModif}).

The pre-existence of a strong hydrodynamic turbulence in the plasma ahead of the shock can also affect the SNR expansion before it reaches Mach and Alfv\'en numbers of order unity.  When the shock velocity drops to values comparable to typical turbulent velocities inside the SB, one may expect large distortions of the shock front.  While such distortions saturate when the shock velocity is large compared to the ambient velocities (as recalled in Sect.~\ref{sec:distortion}), the situation is different when different parts of the shock propagate in fluids with large velocity fluctuations.  If we assume, at zeroth order, that the shock velocity \textit{relative to the local fluid} remains approximately constant, then strong shear of the shock front will start when $V_{\mathrm{SNR}} \sim V_{\mathrm{turb}}$.  Following Bykov and Fleishman  (1992), we can estimate typical turbulent velocities inside SBs of the order of 300 -- 1000~km/s or even more.  This is consistent with our picture of the SB core as a turbulent medium resulting from the interaction of strong stellar winds and SN shocks with dense clumps as well as other shocks, generating numerous secondary shocks.  With such values of $V_{\mathrm{turb}}$, strong fluctuations of the SNR front and of the magnetic field lines attached to it start about $2\,10^{4}$ -- $10^{5}$~years after the explosion (i.e. around or soon after the end of the free expansion phase for the largest turbulent velocities).

For all the above reasons, the evolution of a SN shock inside a superbubble (i.e. a hot, rarefied, inhomogeneous and turbulent medium) is different from that of an isolated SN. Although reliable quantitative estimates would require in-depth studies which are beyond the scope of this paper, we note that all the above-mentioned mechanisms tend to produce strong turbulence and generate MHD waves, turning a significant fraction of the SN explosion energy (which is usually lost in isolated SNRs) into a form which can be available for further particle acceleration. While the first few $10^{4}$~years of the SNR evolution inside a SB should follow the standard scheme (although with a much longer free expansion phase), a series a transition should then occur, following the hierarchy $V_{\mathrm{turb}} \ga V_{\mathrm{Alfven}} \ga c_{s}$, and degrade the shock energy into turbulence, MHD waves and CRs.

As far as energy balance is concerned, it is also interesting to note that not only does a larger fraction of the SN kinetic energy go into turbulence and MHD waves inside a SB than in the free ISM (especially since the radiative phase is never reached), but the stellar wind energy is also feeding the process efficiently (which is not the case for isolated massive stars), and can therefore be used for particle acceleration.

\section{Shock acceleration inside SBs}

Before we turn to the description of a specific SB-acceleration mechanism, with no equivalent in isolated SNRs, let us now investigate the influence of the SB characteristics on the standard SN shock acceleration mechanism, and discuss possible manifestations of collective acceleration effects due to the repeated shocks.

Even though most SN explosions occur inside superbubbles rather than in the free ISM, it could be argued that this does not significantly change the cosmic-ray origin scenario and that diffusive shock acceleration, resulting from the velocity discontinuity at the shock front, should produce essentially identical results wherever the SN shock is located.  Several properties of the SB, however, weaken this argument.

\subsection{Modification of diffusive shock acceleration}
\label{sec:DSAModif}

The efficiency of particle acceleration around strong shocks and the maximum energy, $E_{\mathrm{max}}$, which can be reached, crucially depend on the level of turbulence and the value of the magnetic field.  Since the size of SNRs and the time available for acceleration are limited, large values of $E_{\mathrm{max}}$ require low diffusion coefficients.  A lower limit to the diffusion coefficient along magnetic field lines is provided by the so-called Bohm scaling, where $D_{\mathrm{B}} = \frac{1}{3}vr_{\mathrm{g}}$ and $r_{\mathrm{g}} = \gamma mv/qB$ is the gyroradius of the particle of mass $m$, charge $q = Ze$ and Lorentz factor $\gamma$ in a field of strength $B$.  This gives $D_{\mathrm{B}} \simeq 3.1\,10^{22}\gamma\beta^{2} Z^{-1}B_{\mu\mathrm{G}}^{-1}\,\mathrm{cm}^{2}\mathrm{s}^{-1}$.  To lower this value, and thus increase $E_{\mathrm{max}}$, one needs larger magnetic fields.

In a number of recent studies of diffusive shock acceleration, attention has been turned to the generation of large magnetic fields on both sides of the shock front, by hydrodynamical instabilities and the non-linear amplification by cosmic-rays of the seed magnetic field (e.g. Lucek and Bell, 2000; Berezhko et al., 2003; Ptuskin and Zirakashvili, 2003).  Clearly, such mechanisms should be even more efficient inside superbubbles where strong magnetic fields are present ahead of the shock.  Likewise, the linear damping of the waves in the background plasma, which limits the amplitude of the random magnetic field through ion-neutral collisions, does not occur inside SBs, where the material is fully ionized.  As for the unavoidable non-linear damping through wave-wave interactions, the situation may again be different inside a SB, because the cosmic-rays are not the only source of the wave growth ahead of the shock, and a steady state should be maintained at a higher level than around isolated SNRs, due to the continuous generation of magnetic turbulence in the background.

Although direct measurements of the magnetic fields inside superbubbles are not available, one can estimate that it is indeed larger than in the average ISM, due to the various mechanisms discussed above.  Turbulence generation through direct wind-wind interactions, shock-clump interactions and shock distortion at mildly super-alfv\'enic velocities should be accompanied by MHD wave generation, all the more efficiently that the medium is ionized and the clumps are themselves magnetized (see Sect.~\ref{sec:clumps}).

Assuming equipartition of the mechanical energy released by the massive stars between thermal pressure, turbulence and magnetic fields, one can obtain magnetic fields of the order of 10 -- 20~$\mu$G. Indeed, evaluating $B$ through $P_{\mathrm{SB}} \simeq B^{2}/8\pi$ from Eq.~(\ref{eq:PSB}) gives $B \simeq 10\,\mu$G, while equating $B^{2}/8\pi$ to the total energy density available inside the SB, $\epsilon \simeq L_{\mathrm{OB}}\times t_{\mathrm{SB}}/\mathrm{V}_{\mathrm{SB}}$ (with $L_{\mathrm{OB}}=10^{38}$~erg/s, $t = 10^{7}$~yr and $R_{\mathrm{SB}}$ from Eq.~(\ref{eq:RSB})), gives $B \simeq 20\,\mu$G. A similar estimate was obtained by Bykov and Toptygin (1988, 2001).

Most recent studies of particle acceleration at shock waves also claim magnetic field amplification around the shock fronts, and this is also supported by the multi-wavelength modeling of SNRs. Although the exact mechanism of the field amplification is not yet established, one could expect that CR-wave interactions, field compression and shock-driven instabilities play an important role (e.g. Lucek and Bell 2000). Assuming an amplification factor $\alpha_{\mathrm{B}}$, one can roughly estimate the maximum energy obtained from standard diffusive shock acceleration inside SBs by following Ptuskin and Zirakashvili (2003) and requiring that $D(E) \le 0.1\,V_{\mathrm{SNR}}R_{\mathrm{SNR}}$ at the end of the free-expansion phase (see also Berezhko et al., 1996):
\begin{equation}
E_{\mathrm{max}} \simeq (1.7\,10^{17}\mathrm{eV})\times Z \times\frac{\alpha_{\mathrm{B}}}{20}\times \frac{B_{\mathrm{SB}}}{10\,\mu\mathrm{G}}.
\label{eq:EMaxDSA}
\end{equation}
We see that values of $E_{\mathrm{max}}$ of the order of $Z\times 10^{17}$~eV (as would be required in order to reach the ankle of the CR energy distribution), require values of $\alpha_{\mathrm{B}}$ of the order of 10--20.  This corresponds to enhanced values of the magnetic field at the shock of the order of 100--200~$\mu$G, which does not seem unreasonable compared to what is usually assumed in isolated SNRs (as deduced from X-ray observations, Berezhko and V\"olk, 2004; Ballet et al., 2004), but additional work is needed to give a sensible conclusion.

Another specificity of diffusive shock acceleration inside SBs is related to the presence of a turbulent and magnetized medium ahead of the shock, which can increase the efficiency of particle acceleration.  In isolated SNRs, while efficient turbulence generation is expected downstream, the diffusion of energetic particles ahead of the shock is conditioned to their own ability to generate resonant Alfv\'en waves.  In a superbubble, such waves should pre-exist to some critical level and provide the seed for amplification by the streaming cosmic-rays.  As shown by Lucek and Bell (2000), the corresponding instability leads to the rapid growth of the modes in resonance with the CRs, which can then be scattered efficiently.  This should be made even easier in a magnetized, turbulent medium such as an SB core, resulting in an increase of the acceleration rate at the higher end of the momentum spectrum, where tuned waves usually do not exist and the CRs leak out of the SNR until resonant waves have sufficiently grown. Note however that the MHD turbulence can also have an indirect effect on cosmic ray propagation by acting as a damping mechanism for cosmic-ray generated waves (Farmer and Goldreich, 2004). In that case, the magnetic field amplification could strongly depend on the wavenumber and the efficiency of particle acceleration on the energy range.

Pre-acceleration in the turbulent flow inside the SB should also modify injection, by increasing the number of particles which are energetic enough to \textit{see} the shock discontinuity.  In an isolated SNR, particle injection in the acceleration process is provided by the tail of the thermal distribution (see e.g. Jones and Ellison, 1991; see also Malkov \& V\"olk, 1995, 1998; V\"olk et al., 2003), which limits the fraction of particles flowing through the shock front to be eventually accelerated to about $10^{-4}$ or $10^{-3}$ at most.  In the case of a SNR inside a SB, the situation is quite different, in principle, as virtually all the pre-existing energetic particles passing through the shock will see the discontinuity and be able to gain energy by diffusing back and forth across the shock front.  The resulting re-acceleration will of course be at the expense of the shock energy, and it is expected to affect the energy balance at the shock transition, as well as the global evolution of the SNR. It is then possible that the shock profile adapts to the EP energy flow and increases the size of precursor, so as to limit particle injection.  A situation where the shock is rapidly quenched by the re-acceleration of pre-existing energetic particles can also be envisaged, and it will be investigated in a separate paper (see also the discussion below). 

In summary, the diffusive shock acceleration mechanism is not fundamentally modified inside a SB, but the conditions there are such that i) the maximum energy possibly reached is naturally higher, because the free expansion phase lasts longer and extends to a much larger radius, and because of a pre-existing turbulent magnetic field ahead of the shock, and ii) the injection mechanism is probably more efficient (and perhaps so much that the shock may be quenched by the reacceleration of a high density of pre-existing CRs).

\subsection{Repeated shock acceleration}

We have shown above that SNRs in a SB environment should lead to a very efficient conversion of the explosion energy into cosmic-rays, because of an increased injection efficiency, and also because the shock never becomes radiative and thus a significant fraction of the explosion kinetic energy can be converted into turbulence and MHD waves inside the SB, which in turn provide an additional acceleration mechanism (see Sect.~\ref{sec:turbulentAcc}).  In this section, we investigate multiple shock acceleration effects, as a result of repeated SN explosion in the SB.

Multiple shock acceleration in the context of SBs has been discussed by Klepach et al. (2000). In their model a large number of strong spherical SN shocks must be simultaneously present in the volume of interest. Such a model would require an extreme SN rate in SB, because of the lifetime of a SN blast wave is of the same order as the time scale between two explosions ($\sim 3\,10^{5}$~yr). Thus the number of coexisting primary SN shocks inside a SB must be small ($\la 2$), unless powerful starburst region, which is not frequent in the Milky Way.

However, \textit{repeated shock acceleration} is quite possible, and must actually occur for relatively low-energy particles. The situation can be straightforwardly described by remarking that since the individual SN shocks become subsonic well inside the SB, \textit{all the EPs} accelerated by diffusive shock acceleration (DSA) at the shock will be released inside the SB and diffuse from there out of the system. Now if the time required for them to leave the SB is larger than the typical time between two SN explosions, they may be overcome by a subsequent shock, and thus be injected into a new DSA episode.

\subsubsection{A toy model}

If one could neglect all other acceleration processes of the particles between two successive shocks (but see Sect.~\ref{sec:turbulentAcc}), the effect of such repeated shock acceleration could be estimated straightforwardly in the test-particle limit, by simply applying several times the 'transfer operator', $\mathcal{T}$, of one shock.  The latter is well know from standard planar DSA theory (e.g. Blandford and Ostriker, 1978; 1980), and can be expressed very simply through a change of variable:
\begin{equation}
\mathcal{T}\circ f_{\mathrm{in}} = xp^{-x}\int_{0}^{p}p^{\prime x-1} f_{\mathrm{in}}(p^{\prime})\mathrm{d}p^{\prime} = \int_{0}^{1}f_{\mathrm{in}}(pu^{1/x})\mathrm{d}u,
\label{eq:transferOperator}
\end{equation}
where $f_{\mathrm{in}}$ is the incoming EP distribution function, and $x = 3r/(r-1)$ is the standard power-law index found in test-particle DSA theory, for a shock with compression ratio $r$. When applied to an initial distribution function, far upstream, given by $f_{\mathrm{in}} = (n_{0}/4\pi p_{0}^{2})\delta(p-p_{0})$ (monoenergetic `injection'), one obtains the well-known results:
\begin{equation}
f_{1} = \frac{n_{0}}{4\pi p_{0}^{3}}\times x \left(\frac{p}{p_{0}}\right)^{-x}\times \mathrm{H}(p-p_{0}),
\label{eq:f1}
\end{equation}
where $\mathrm{H}(x)$ is the Heaviside function.

Analytical iteration of the transfer operator is possible, and one can thus obtain the distribution function of the EPs after $n$ shocks (i.e. $n$ iterations of $\mathcal{T}$), assuming that test-particle is still valid (c.f. White 1985; Achterberg 1990):
\begin{equation}
f_{n}(p) = \mathcal{T}^{n}\circ f_{\mathrm{in}} = \frac{x^{n-1}}{(n-1)!}\left[\log(p/p_{0})\right]^{n-1} f_{1}(p).
\label{eq:fn}
\end{equation}

The above formula includes the compression factor through the shock, so that the EP number density (obtained by integration over $p$) is $n_{0}r^{n}$.  A proper account the necessary decompression of the shocked gas between two DSA episodes, without which the SB would actually be shrinking, should also affect the EP momentum distribution.  If the EPs are coupled to the hot gas behind the shock after they have left the acceleration process (but still in the compressed region), they should experience adiabatic losses corresponding to a dilation inverse of the shock compression.  In such a process, the particles of momentum $p$ end up with momentum $p\times r^{-n/3}$, and the EP distribution function after $n$ shock crossings should actually be written $f_{n}^{\prime}(p) = f_{n}(pr^{n/3})$, with the above expression for $f_{n}$.  If on the other hand the EPs integrate the general flow inside the SB without significant energy losses, the effective distribution function to be considered after $n$ DSA episodes should simply write $\tilde{f}_{n}(p) = f_{n}(p)/r^{n}$.

At a given time of the repeated shock acceleration process, EPs having passed through various numbers of shocks coexist inside the SB. The effective distribution function is thus given by the sum of $\tilde{f}_{n}$ functions, with $n$ ranging from 1 to $N$, the maximum number of shocks seen by one particle, which depends on the age of the SB (and explosion rate).  The sum should be weighted by the probability that a particle has remained inside the SB long enough to be (re-)accelerated by the corresponding number of shocks.  If $\mathrm{P}_{\mathrm{esc}}$ is the escape probability and we write $q = 1 - \mathrm{P}_{\mathrm{esc}}$, one obtains (using $\tilde{f}_{n}$ functions for simplicity):
\begin{equation}
F_{N}(p) = \sum_{n=1}^{N}q^{n}\frac{f_{n}(p)}{r^{n}}.
\label{eq:sumfn}
\end{equation}
In the limit of large $N$, this sum tends towards:
\begin{equation}
F_{\infty}(p) = \frac{n_{0}}{4\pi p_{0}^3}\frac{qx}{r} \left(\frac{p}{p_{0}}\right)^{-3-3P_{\mathrm{esc}}/(r-1)},
\label{eq:FInfty}
\end{equation}
for $p \ge p_{0}$, where one recognizes a generalization of the well-know result that multiple shock acceleration leads to a hard spectrum in $p^{-3}$ (instead of $p^{-4}$) if there is no escape.  For finite values of $N$, $F_{N}(p)$ also shows a $p^{-3}$ behaviour at low energy (where the truncated sum is very close to the infinite one due to rapid decrease of higher order terms), up to higher and higher energies when $N$ increases.

Obviously, the above is nothing but a toy model, and the obtained solution is unrealistic in several respects.  First of all, it was obtained in the test-particle approximation (i.e. without retroaction of the EPs on the shock structure), while we have argued that the high density of EPs inside SBs should significantly modify the flow (see also below).  In addition, we have neglected all other type of acceleration, such as turbulent acceleration which may be the dominant one, as we argue below. Further acceleration of the particles between two shocks should thus modify the resulting spectrum.  Finally, the probability that an EP reaches the shell of the SB and/or escapes before another shock arrives is an energy-dependent function, and depends also on the sequence of SN explosions and on the EP initial position. Nevertheless, we use the simple model above to argue that repeated shock acceleration must occur inside SBs, at least up to energies such that $\tau_{\mathrm{esc}}(E) \la \Delta t_{\mathrm{SN}}$. For these particles, a hardening of the spectrum is to be expected, and be it only for that reason the acceleration process cannot be considered as identical to what is encountered in isolated SNRs (a fortiori if another mechanism actually dominates).

\subsubsection{Maximum energy of repeatedly accelerated particles}

Let us now estimate the maximum energy of the EPs which indeed encounter several SN shocks before they leave the SB. The typical `escape time' is given by $\tau_{\mathrm{esc}} \sim R^{2}/2D$, where $D(E)$ is the average diffusion coefficient in the SB. As shown by Casse et al.~(2002), the variation law of the diffusion coefficient with rigidity depends on the ratio, $\rho = r_{\mathrm{g}}/\lambda_{\mathrm{max}}$, of the EP gyroradius to the principal length scale of the turbulence.  For the low-energy particles considered here, $\rho \ll 1$, and the Bohm diffusion regime is not reached.  Given the expected high level of turbulence, one can assume roughly isotropic diffusion, with a diffusion coefficient of the order of:
\begin{equation}
D(E) = \frac{1}{3}\lambda_{\mathrm{max}}c\, \eta_{\mathrm{T}}^{-1}\rho^{2-\beta},
\label{eq:D}
\end{equation}
where a power-law turbulent spectrum of index $\beta$ was assumed, $S(k)\propto\eta_{\mathrm{T}}(k\lambda_{\mathrm{max}})^{-\beta}$, and $\eta_{\mathrm{T}} = \left<\delta B^{2}\right>/(\left<\delta B^{2}\right> + B^{2})$ is probably close to~1, as the turbulent field should dominate (see above).

With the above assumptions, the diffusion coefficient is estimated for a turbulent length scale of the order of the typical distance between massive stars, $R_{\star}$, given in Eq.~(\ref{eq:Rstar}).  For the characteristics of our typical OB association, $R_{\star} \simeq 6$~pc, and one obtains:
\begin{equation}
D(E) \simeq (1.0\,10^{27}\,\mathrm{cm}^{2}\mathrm{s}^{-1}) \left(\frac{\lambda_{\mathrm{max}}}{6\,\mathrm{pc}}\right)^{2/3} \hspace{-5pt}\eta_{\mathrm{T}}^{-1} E_{\mathrm{GeV}}^{1/3}B_{\mu\mathrm{G}}^{-1/3}.
\label{eq:DNum}
\end{equation}

The maximum energy for repeated shock acceleration, $E_{\mathrm{rsa}}$, is then obtained from the condition $D(E_{\mathrm{rsa}}) \simeq R^{2}/2\Delta t_{\mathrm{SN}}\sim 1.1\,10^{28}\,\mathrm{cm}^{2}\mathrm{s}^{-1}$.  Taking the maximum SN extension $R \simeq R_{\mathrm{sub}}$ (cf.  Eq.~\ref{eq:RSub}) and $\Delta t_{\mathrm{SN}} \simeq 3\,10^{5}$~yr, one finds\footnote{This value of the limiting diffusion coefficient is larger than the value obtained for turbulent diffusion under the SB conditions, i.e. $v_{\mathrm{turb}}\la 10^{3}$~km/s and $l_{0} \sim R_{\star} \sim 6$~pc.  Therefore, it is legitimate to use the non-turbulent expression, Eq.~(\ref{eq:D}), for the order of magnitude calculation.}:
\begin{equation}
E_{\mathrm{rsa}} \simeq 11\,\mathrm{TeV}\times \eta_{\mathrm{T}}^{3} \left(\frac{B_{\mathrm{SB}}}{10\,\mu\mathrm{G}}\right) \left(\frac{\lambda_{\mathrm{max}}}{6\,\mathrm{pc}}\right)^{-2} \hspace{-5pt}.
\label{eq:Ersa}
\end{equation}
where we have used $10\,\mu$G as a fiducial value of the magnetic field inside SBs, which corresponds to a factor $\sim 2$ less than the equipartition value.

In the absence of any other mechanism (but see Sect.~\ref{sec:turbulentAcc}) and if the shocks remain unmodified, this energy would typically mark a smooth transition between a $p^{-3}$ and a $p^{-4}$ EP spectrum (assuming strong shocks with a compression ratio $r=4$).

Finally, before we turn to a different acceleration mechanism specific to superbubbles, let us comment briefly on the question of shock modification.

\subsubsection{Energy crisis and shock modification}

In applying the above toy model for repeated shock acceleration, we assumed that the test-particle approximation could be used.  As is well know from DSA theory, such an approximation cannot hold if the acceleration is efficient enough and a significant fraction of the shock energy is imparted to the EPs.  In this case, the EPs influence the shock dynamics, and the compression ratio across the discontinuity.  This in turn modifies the EP distribution function non linearly.  In a SB, the situation is aggravated because of the repeated shock acceleration effect.  Indeed, when a SN shock travels in the SB medium, a large number of pre-existing EPs are injected into the DSA process, in addition to the usual high-energy tail of the shocked gas thermal distribution.  The EPs of a previous generation which have not diffused away from the region swept up by the new shock (i.e. with energies lower than $E_{\mathrm{rsa}}$) have large enough gyroradii to see the shock discontinuity, and thus will gain energy from the velocity difference by diffuse back and forth across the shock.  Now this energy gain will of course be at the expense of the shock energy. So it is interesting to estimate the amount of energy involved.

By essence, all the EPs do not gain the same amount of energy, as it depends on the number of shock crossings and as well as the crossing angles.  However, placing ourselves in the test-particle approximation, as in the toy model discussed above, it is easy to estimate the average energy gain per particle.  For a particle `injected' in the shock with momentum $p_{0}$, the average energy at the end of the DSA mechanism is obtained by integrating $f_{1}(p)\times pc$ (for relativistic particles), where $f_{1}(p)$ is the distribution function given by Eq.~(\ref{eq:f1}). Dividing by $p_{0}c$, one gets the mean energy amplification factor: $E_{1}/E_{0} = \ln (p_{\mathrm{max}}/p_{0})$, if $x = 4$, or $E_{1}/E_{0} = (x-3)/(x-4)$, if $x > 4$.

For standard, un-modified strong shocks (compression ratio of 4), the spectral index is $x = 4$ and the energy gain per particle is quite large.  For an EP of initial energy $E_{0} = 1$~GeV, and even for low maximum energy of the order of 1~TeV, the energy gain is by a factor of $\ln 100 \simeq 7$.  So to be rather conservative, let us assume that a first SN shock has given 10\% of its energy to CRs, and that a fraction $1/3$ of this energy is in CRs of sufficiently low-energy to remain inside the SB until a new shock arrives.  Then all these CRs will be re-accelerated to an average energy higher by a factor of 7 or even larger (for higher values of $p_{\mathrm{max}}$), which will cost about 20\% of the new shock's energy.  This energy budget will then keep on increasing with the number of shocks exploding inside the SB. Note that the above estimate is actually very conservative, as the value of $p_{\mathrm{max}}$ should be much higher than 1~TeV/c.

If nothing could modify the situation, the result of this energy crisis would be that the shocks propagating inside an already active SB quickly exhaust themselves by re-accelerating EPs from previous generations.  Before that, of course, the EPs will start to play a major role in the shock dynamics (and MHD wave generation).  From the above estimate, it is clear that a steeper spectrum (larger value of $x$, i.e. smaller compression ratio, or weaker shock) can lower the energy ratio $E_{1}/E_{0}$.  The lower compression ratio could also be obtained through a broadening of the shock, in so far as EPs can only be shock accelerated if their gyroradius is larger than the shock thickness.  In other words, the non-linear effect will work in such a way that the injection of previously existing EPs will be reduced.  But in that case, of course, the shock will be a poor accelerator of the ambient thermal material.  In conclusion, the resolution of the above-mentioned energy crisis in real SBs deserves a more detailed analysis, but whatever it may be, it is another important difference between isolated SNRs and SN shocks expanding inside SBs.

\section{Turbulent acceleration inside SBs}
\label{sec:turbulentAcc}

Having discussed the collective effects associated with repeated shock acceleration of relatively low-energy particles, and how the standard DSA mechanism should be modified inside SBs, we now turn to the description of a specific mechanism with no equivalent in isolated SNRs, and which may be responsible for most of the energy transfer from SN and stellar wind energy to energetic particles. This mechanism has been studied by Bykov and Toptygin (1987, 1990, 2001), Bykov and Fleishman (1992), Bykov (1995, 2001), and we only give here an outline of its main features.

The idea is to describe particle interaction with a complex ensemble of multiple MHD shocks and large scale motions produced by the interaction of strong (primary) SN shocks with inhomogeneites like the shells of ambient matter swept up by stellar winds or cloud fragments (see Sect.~\ref{sec:clumps}). The general kinetic theory is applied in this context and the effective kinetic equation satisfied by the EP distribution function is derived for the velocity field of a superbubble described statistically, taking into account the ensemble of multiple shocks and the associated long-wavelength MHD waves in the low-density, highly turbulent and magnetized plasma which fills the SB.

This shock ensemble is typically dominated by weak shocks and described by a number of cross-correlation functions. The MHD shocks produce an intermittent distribution of accelerated particles with strong  fluctuations in the low energy part of the spectrum. According to the model, this part of the distribution function could contain a substantial part of the energy released in SBs. The linear treatment of the acceleration indicates that the energy conversion is very efficient indeed, so that the retroaction of the accelerated particles must be considered. Bykov (2001) then developed a non-linear approach of the SB accelertation mechanism, describing the link between the EPs and the MHD wave ensemble. It was shown that 20--40$\%$ of the kinetic power released in the SB can be transferred to low-energy particles on a time scale shorter than $\sim 10^6$ years, and a time-dependent spectrum of accelerated particles could be obtained.

Interestingly enough, the time asymptotic distribution function is found to be a power-law momentum distribution, with a logarithmic index in the range $4 \leq x \leq 5$. The index is close to 5 if the gas pressure is dominated by the non-relativistic component, and it approaches 4 in the case of a relativistic gas pressure. Note that the model assumed the presence of small-scale MHD fluctuations of wavelengths below particles mean free-path. This is supported by recent 3D simulations showing the development of magnetic field fluctuation spectra due to large scale motions of a highly conducting plasma (e.g. Biskamp, 2003).

An important advantage of the weak shock acceleration scenario is that the efficiency of particle acceleration is then higher than that of ambient gas heating. This is generally true for a shock with sonic Mach number $M \lsim 1 + \beta^{-1}$ (Bykov and Toptygin 1987). Thus, in a magnetized system with $\beta\equiv 8\pi P/B^2 \lsim 1$, even shocks with $M \gsim 2$ transfer most of their kinetic energy to the non-thermal particles.

In a recent analysis of the observed energy budget of superbubble DEM L 192 (or N 51D), in the Large Magellanic Cloud, Cooper et. al (2004) found a discrepancy between the stored thermal and kinetic energies, representing only (6 $\pm$ 2)$\times 10^{51}$ ergs, and the injected kinetic energy estimated to be (18 $\pm$ 5)$\times 10^{51}$ ergs. A natural solution to this apparent energy crisis could be that a substantial amount of the injected energy has been converted into magnetic fields and non-thermal particles. The conversion efficiency required to solve the problem is of the order of that expected within the SB acceleration model.

Regular and stochastic magnetic fields are governing the maximal energies of accelerated CRs. As discussed above, magnetic fields of the order of 10 -- 20~$\mu$G could be common inside SBs. As far as individual shocks are concerned, an estimate of $E_{\mathrm{max}}$ was given in Eq.~(\ref{eq:EMaxDSA}), which could reach the ankle region if efficient field amplification operates around the shock, and if the Bohm diffusion regime holds. Further away from the primary strong shocks, such a regime probably does not hold. In an alternative model of CR diffusion inside a superbubble, the particles are scattered by multiple secondary weak shocks. This is the typical situation of EPs experiencing turbulent SB acceleration in the intervals between two passages of major SN strong shocks. For such a mechanism, Bykov and Toptygin (2001) found a maximum energy of the SB-accelerated EPs around $E_{\mathrm{max}} \sim 10^{17}$~eV, compatible with the highest energy Galactic CRs. They also made a prediction for the CR composition above the knee, and showed that a thorough measurement of the mean CR atomic weight as a function of energy (i.e. $<\ln A(E)>$) could test the models.

\section{Conclusion}

We have reviewed the possible collective effects of particle acceleration associated with the explosion of numerous SNe in a limited region of space and on a short timescale.  We discussed several aspects of the problem, each of which, on its own, gives evidence that particle acceleration inside SBs acts in a different way from the standard diffusive shock acceleration mechanism prevailing at isolated SNRs.  This is our main conclusion.  We did not try and solve the complicated problem of EP acceleration in SBs, but rather showed that it deserves detailed investigation, especially since, as we reminded, most of the energy released in the ISM by massive stars is injected inside SBs, and therefore SBs should be considered as the most probable source of CRs.

Among the main conclusions reached in this paper, we have shown that the massive stars in OB associations are usually close enough to one another not only for their wind bubbles to interact and merge (forming the SB), but also for their unshocked wind material to expend up to distances larger than half the mean distance between OB stars, so that they can directly interact (or terminate each others).  In the interaction region, efficient generation of strong turbulence and MHD waves should occur, maintaining conditions propitious for turbulent particle acceleration.  We have also shown that cloudlets or clumps of higher density material (most probably magnetized) should be present inside the SB, providing additional seeds for turbulence and MHD waves through their interaction with the primary and secondary shocks induced by the intense stellar activity.

The global behaviour of SN shocks inside SBs should be roughly similar to what is observed in the standard ISM. However, a few significant differences should manifest.  While the shock distortions in the ambient turbulent medium should be saturated during the first few tens of kiloyears, sound waves and MHD waves should then be produced with high efficiency when the shock becomes mildly super-alfv\'enic and supersonic.  Most significantly, we have shown that the alfv\'enic and sonic transitions occur i) before the shock becomes radiative, so that no energy is lost from the system (contrary to the case of isolated SNRs) and ii) well inside the SB, so that the remaining energy is released in the hot interior, and is thus available for further particle acceleration.  In other words, not only do the Galactic SNe occur most often inside superbubbles, but they should also be more efficient in accelerating particles there than in the rest of the ISM, as follows from energy balance considerations.

In addition, we have shown that the lowest energy particles (possibly up to the TeV range) will experience repeated shock acceleration, as the EPs accelerated at one SN shock do not have time to diffuse out of the SB before the next SN shock sweeps the SB interior.  This has several interesting consequences.  First, a hardening of the spectrum can be expected at low energy (as is common in multiple shock acceleration).  Second, the presence of previously accelerated particles in the upstream region of a SN supersonic flow can in principle make injection (into the acceleration process) very efficient.  All the EPs with a gyroradius much larger than the shock thickness will `see' the shock discontinuity, and experience diffusive shock acceleration. For this reason, an energy crisis is likely to occur, where the EP re-acceleration quickly exhausts the SN shock energy.  To avoid this, non-linear effects are expected to modify the flow and/or lower the injection efficiency, so that diffusive shock acceleration may turn out to be quite different inside and outside a superbubble.  In the above process, some fraction of the shock energy can be transferred to magnetic fields, thereby feeding a different acceleration mechanism, specific to SBs.

Indeed, we have argued that various mechanisms (from direct wind-wind interactions to shock-cloud interactions and shock distortion at late times) maintain a high level of turbulence and magnetic inhomogeneities in (at least part of) the SB interior -- which we can refer to as its core, and that turbulent acceleration should be very efficient in this core.  This is a result of standard kinetic theory, whose application to a SB environment has been extensively studied for more than decade (e.g. Bykov and Toptygin 1990).  The result of the linear theory is that particle acceleration should be so efficient that the retroaction of the EPs on the flow and MHD waves must be included.  First attempts to do so in a stochastic approach have shown that  power-law EP distribution functions can be expected quite naturally, although the index of the power-law depends on the details of the injection processes (either from strong shock acceleration, fast moving knots, or resonant particle injection). However that may be, power-laws steeper than $E^{-2}$ seem common to obtain inside SBs, which may be seen as a interesting result in the context of the cosmic-ray source theory.  More generally, the ideas discussed above have some consequences for the GCR problematics as well as for non thermal astronomy.  These are discussed in detail in two accompanying papers.

From a general point of view, it is interesting to note that, contrary to the case of isolated SNRs, SB environments offer a unique opportunity to use not only the SN explosion energy, but also the energy of the strong stellar winds.  In SB cores, the latter naturally feeds the turbulent acceleration mechanism by providing both secondary shocks and MHD waves, while the termination shocks of isolated massive stars do not seem to be efficient particle accelerators, probably because of the expected low value of the local magnetic field upstream (i.e. in the wind itself, far from the star).

In this paper, we have only considered "standard" superbubbles, resulting from the activity of typical OB associations in the Galaxy.  One should also think, however, of the huge OB clusters which are found in the center of most galaxies, including ours (e.g. Figer, 2003).  These can be seen as on-going star bursts, with huge stellar densities (and particularly flat IMFs!), where the direct wind-wind interactions must be extremely important.  In such regions, the SB acceleration process described above should be particularly efficient, and impossible to analyze as a mere succession of isolated SNR acceleration processes. However, the corresponding environment is probably harder to control, as strong gas expansion (and possibly galactic winds) may be generated in such bursts.  For this reason, we have limited our study to the observationally better-defined SBs, although the contribution of the central region of the Galaxy to the observed CR flux may also be important.

Finally, it should be noted that efficient particle acceleration inside superbubbles may have consequences on the phenomenology of the SBs themselves.  In particular, if a large fraction of the internal energy is in relativistic particles, the effective adiabatic index in the SB interior may be smaller than usually assumed, which would modify the dynamics of the SB. Energy leakage through high-energy particles could also affect the SB evolution, and maybe help reconciling observations and theory. This will be addressed elsewhere.

\begin{acknowledgements}
We wish to thank Don Ellison and Yves Gallant for stimulating discussions, as well as Luke Drury for inviting three of the authors at the Dublin Institute for Advanced Studies, where this work was initiated. The authors also acknowledge support from the GDR PCHE program of the french National Scientific Research Center, CNRS.  A.M.B. was partially supported by RBRF grant 03-02-17433.
\end{acknowledgements}

\end{document}